\documentclass[12pt,letter]{article}

\usepackage{graphicx, epsfig, color,cite}
\usepackage{amsmath}
\usepackage{amssymb}
\usepackage{subfigure}

\def\eslt{E_T^{\rm miss}}
\def\delew{\Delta_{\rm EW}}

\def\to{\rightarrow}

\def\bi{\begin{itemize}}
\def\ei{\end{itemize}}

\def\sps1ap{SPS1a$^\prime$}
\def\c1p{C1$^\prime$}

\def\tb{\tilde b}

\def\tst{\tilde t}

\def\tg{\tilde g}

\def\tq{\tilde q}
\def\tw{\widetilde W}
\def\tz{\widetilde Z}

\def\be{\begin{equation}}  
\def\ee{\end{equation}}  
\def\bea{\begin{eqnarray}}  
\def\eea{\end{eqnarray}}  
\def\beas{\begin{eqnarray*}}  
\def\eeas{\end{eqnarray*}}  
\newcommand\prd[3]{{\it Phys.\ Rev.\ }{\bf D #1} (#2) #3}

\newcommand\plb[3]{{\it Phys.\ Lett.\ }{\bf B #1} (#2) #3}
\newcommand\jhep[3]{{\it J. High Energy Phys.\ }{\bf #1} (#2) #3}

\newcommand\npb[3]{{\it Nucl.\ Phys.\ }{\bf B #1} (#2) #3}

\newcommand{\hepph}[1]{hep-ph/#1}

\begin{document}
\begin{titlepage}
\begin{flushright}
OU-HEP-150430 \\
FTPI-MINN-15/23 \\
IPMU15-0070
\end{flushright}

\vspace{0.5cm}
\begin{center}
{\Large \bf Prospects for Higgs coupling measurements\\
in SUSY with radiatively-driven naturalness
}\\ 
\vspace{1.2cm} \renewcommand{\thefootnote}{\fnsymbol{footnote}}
{\large Kyu Jung Bae$^1$\footnote[1]{Email: bae@nhn.ou.edu},
Howard Baer$^1$\footnote[2]{Email: baer@nhn.ou.edu },
Natsumi Nagata$^{2,3}$\footnote[3]{Email: natsumi.nagata@ipmu.jp}
and Hasan Serce$^1$\footnote[4]{Email: serce@ou.edu}
}\\ 
\vspace{1.2cm} \renewcommand{\thefootnote}{\arabic{footnote}}
{\it 
$^1$Dept. of Physics and Astronomy,
University of Oklahoma, Norman, OK 73019, USA \\[3pt]
}
{\it 
$^2$William I. Fine Theoretical Physics Institute, School of Physics and
 Astronomy, \\
University of Minnesota, Minneapolis, MN 55455, USA \\[3pt]
}
{\it 
$^3$Kavli IPMU (WPI), UTIAS, University of Tokyo, Kashiwa, Chiba
 277-8583, Japan\\ 
}

\end{center}

\vspace{0.5cm}
\begin{abstract}
\noindent 
In the post-LHC8 world-- where a Standard Model-like Higgs boson
has been established but there is no sign of supersymmetry (SUSY)--
the detailed profiling of the Higgs boson properties has emerged as
an important road towards discovery of new physics.
We present calculations of the expected deviations in Higgs boson
couplings $\kappa_{\tau ,b}$, $\kappa_t$, $\kappa_{W,Z}$, $\kappa_g$ and
$\kappa_\gamma$ versus the naturalness measure $\delew$. 
Low values of $\delew\sim 10-30$ give rise to a natural Little Hierarchy
characterized by light higgsinos with a mass of $\mu\sim m_Z$ while top
squarks are highly mixed but lie in the several TeV range.
For such models with radiatively-driven naturalness, one expects the
Higgs boson $h$ to look very SM-like although deviations can occur. 
The more promising road to SUSY discovery requires direct higgsino pair 
production at a high energy $e^+e^-$ collider operating with the
 center-of-mass energy
$\sqrt{s}>2\mu\sim \sqrt{2\delew}m_Z$.

\vspace*{0.8cm}

\end{abstract}

\end{titlepage}

\section{Introduction}
\label{sec:intro}

The recent discovery of a Standard Model (SM)-like Higgs boson
\cite{atlas_h,cms_h} with mass $m_h=125.09\pm
0.24$~GeV \cite{Aad:2015zhl} (ATLAS/CMS combined values)  
is a triumph of contemporary physics in that it provides the 
first hard evidence for the existence of fundamental scalar fields. 
Theoretically, such  spinless fields are hard 
to comprehend due to unstable quadratic quantum corrections 
to their mass value \cite{susskind}. If nature is supersymmetric (SUSY), 
then the unwanted divergences are cancelled to all orders in 
perturbation theory thus allowing for a naturally occurring 
Higgs boson \cite{wittenkaul}. Yet, so far, no sign of softly-broken
weak scale SUSY has appeared at LHC \cite{atlas_s,cms_s}. 
The growing mass gap between the weak scale-- 
as typified by the $W$, $Z$ and Higgs boson masses $\sim 100$~GeV--
and the sparticle mass scale $m({\rm sparticle})\gtrsim 1-2$~TeV has led to
the re-emergence of the naturalness question: this time involving
log rather than quadratic divergences.

Recent evaluations of supersymmetric models with radiatively-driven 
naturalness
(RNS for radiatively-driven natural SUSY\cite{rns}) find that for a value of
$\delew<10$ (30), then the gluino mass is bounded from above by
$m_{\tg}\lesssim 2.5$~TeV (5~TeV). These $\delew$ values correspond to
$\delew^{-1}=10\%$ (3.3\%) fine-tuning respectively. In contrast,
the $5\sigma$ reach of LHC13 (LHC with $\sqrt{s}=13$~TeV) for gluino
pair production is estimated to be $m_{\tg}\sim 1.6$~TeV (100~fb$^{-1}$)
and 1.9~TeV (1000~fb$^{-1}$) \cite{andre,lhc}. In RNS models with
gaugino mass unification, this reach can be extended via the same-sign
diboson signature arising from wino pair production to equivalent values
of $m_{\tg}\sim 2.4$~TeV for 1000~fb$^{-1}$ \cite{lhc,ssdb}.
The upshot is that LHC13 may or may not have sufficient
energy/luminosity to fully probe the entire parameter space of natural
SUSY.\footnote{Prospects for LHC13 indirect searches for RNS via initial
state radiation off of higgsino pair production reactions (monojet
signal) seem pessimistic\cite{chan}. Allowing for monojet radiation off of
$\tz_1\tz_2$ production, $\tz_2\to\tz_1\ell^+\ell^-$ may allow probes of
the higgsino mass $\mu$ up to $\sim 200$~GeV assuming $\sim
1000$~fb$^{-1}$  of integrated luminosity \cite{jll}.}

Without a guaranteed path towards SUSY discovery, other alternatives
have been explored. Many recent investigations promote Higgs boson
profiling as a probe for physics beyond the SM. Since the
Higgs boson $h$ is now discovered, the goal is to measure every possible
property of $h$ to see if they maintain consistency with the SM or
produce deviations which might point to new physics. These quantities
include: the mass and width of the Higgs boson, its spin (which is
essentially already determined to be spin-0 \cite{Khachatryan:2014kca})
and its coupling strengths to various SM and non-SM decay modes. The
coupling strengths $\kappa_i$ are usually parametrized in terms of the
SM values. Thus, $\kappa_b\equiv
g_{hb\bar{b}}/g_{hb\bar{b}}(\text{SM})$, $\kappa_Z\equiv
g_{hZZ}/g_{hZZ}(\text{SM})$, {\it etc.}. Evidence for beyond the SM (BSM) physics would
then occur from the measurement of one or more $\kappa_i$ values
($i=\tau ,b,t,Z,W,g,\gamma$ ) to significantly differ (by several error
bars) from the SM value of 1. 

The capability of various accelerator options to measure the $\kappa_i$
values has been analyzed \cite{Han:2013kya,Peskin:2013xra} and tabulated
in Ref's.~\cite{Dawson:2013bba, Moortgat-Pick:2015yla}. In fact, early
data from LHC8 seemed to indicate an enhancement in the Higgs to
diphoton coupling which could have been construed as requiring new
TeV-scale charged particles that circulate in the $h\gamma\gamma$ loop
\cite{Carena:2011aa,ArkaniHamed:2012kq,Djouadi:2013lra}. However, the current profile
of the Higgs boson is consistent with SM predictions: {\it i.e.}, at
present the Higgs boson appears to be ``the'' SM Higgs boson as no credible
deviations from the SM have been found. As more data accrue from various
collider options, the error bars on the various Higgs observables will
tighten, and may reveal physics beyond the
SM \cite{Cahill-Rowley:2014wba,Endo:2015oia, Bhattacherjee:2015sga}. 

A particularly interesting scenario which merits investigation is that
of natural supersymmetry. In fact, several previous works have already
investigated this case: Ref's.~\cite{kmm,fr,hm,frw}. These papers all
investigated models with light third generation squarks which are a
consequence of minimizing the ``large log'' contribution to the Higgs
boson mass: $\delta m_h^2\sim
-\frac{3f_t^2}{8\pi^2}(m_{Q_3}^2+m_{U_3}^2+A_t^2)\ln\left(\Lambda/m_{\rm
SUSY} \right) $, where $f_t$ is the top Yukawa coupling, $\Lambda$ is the
cutoff scale, $m_{\text{SUSY}}^2=m_{\tst_1}m_{\tst_2}$ is the SUSY-breaking scale, 
and $m_{Q_3}^2$, $m_{U_3}^2$, and $A_t$ denote the soft masses and the $A$-term for
stops, respectively. The validity of this measure has been challenged in
Ref's.~\cite{comp,seige} in that it sets to zero additional dependent
contributions which lead to large cancellations. Alternatively, it is argued
that the correct measure is $\delew$ which instead requires 1. that the
SUSY $\mu$ term is comparable to the weak scale ($m_{\rm weak}\sim
100$~GeV as typified by the $W$, $Z$ and $h$ masses), 2. that
$m_{H_u}^2$ is driven radiatively to negative values of magnitude
comparable to $m_{\rm weak}$ and 3. that radiative corrections
to the weak scale effective potential\footnote{$\Sigma_u^u$ and $\Sigma^d_d$
are given in the Appendix of Ref.~\cite{rns}.} (which determines the electroweak
vacuum expectation values (VEVs) and hence the $Z$-boson mass $m_Z$) are
comparable or less than $m_{\rm weak}$. This latter condition is met for
highly mixed but TeV-scale top-squarks, 
{\it i.e.}, much heavier than values expected from large log minimization. 
Meanwhile, the first of these conditions implies a
spectrum of four higgsino-like states $\tz_{1,2},\tw_1^\pm$ with the
$\tz_1$ as a higgsino-like lightest SUSY particle (LSP) and 
candidate for dark matter.  
If naturalness is required as well in the QCD sector, then the axion
solution to the strong CP problem is invoked \cite{Peccei:1977hh}. The
SUSY DFSZ axion model \cite{Dine:1981rt} not only tames the strong CP
problem but also provides an elegant solution to the SUSY $\mu$
problem. In this class of models, the apparent Little Hierarchy as typified
by $\mu\ll m(\text{sparticle})$ can be naturally generated via a radiative
breakdown of Peccei-Quinn symmetry\cite{msy,bbs}. 

While we agree with the assessment of Ref.~\cite{craig} that unnatural
SUSY is likely to be wrong SUSY, we would disagree with the assessment
that the minimal supersymmetric Standard Model (MSSM) is at present
fine-tuned over all of parameter space. 
While many SUSY models are indeed fine-tuned under $\delew$\cite{seige}, 
the class of SUSY models with radiatively-driven
naturalness (RNS) remain highly natural. The reason is that the current
experimental limits on the SUSY $\mu$ parameter arise from negative
searches for chargino pair production at LEP2: $m_{\tw_1}>103.5$~GeV. 
Roughly speaking then, $\mu$ is also $\gtrsim 100$~GeV. This value is
quite close to the value of $m_Z$ so that we can interpret the fact
that $m_W, m_Z$ and $m_h$ are all clustered near 100~GeV as a
consequence of $\mu^2$, $m_{H_u}^2(\text{weak})$ and $\Sigma_u^u$ all
being comparable to-- or lighter than- (100~GeV)$^2$.
Meanwhile, the other soft SUSY breaking parameters can be much heavier,
as is indicated by LHC sparticle search limits and radiative corrections
to $m_h$. 

As mentioned above, the class of RNS models predict light higgsino-like
states around the electroweak scale. In addition, there may also be
electroweak gauginos and/or heavy Higgs bosons below the TeV scale,
which are currently less constrained as they are un-colored
particles. The presence of such particles can in principle modify the Higgs
couplings. For instance, the chargino loop contribution can
alter the $h\gamma\gamma$ coupling $g_{h\gamma\gamma}$ if the chargino state
has a sizable wino component. Also, if heavy Higgs bosons have
relatively small masses, the lightest Higgs-boson couplings deviate from
those in the case of decoupling limit, {\it i.e.}, the SM Higgs
ones. The precise measurements of the Higgs couplings, therefore, may
provide a way of probing the RNS scenario indirectly. Since forthcoming
collider experiments can offer significantly improved sensitivities, it
is quite important to investigate whether these experiments can actually
observe any deviations in Higgs couplings in the case of RNS models. 

In this paper, we calculate the deviations to the Higgs boson couplings
$\kappa_{\tau,b,t,W,Z,g,\gamma}$ in supersymmetric models with low
$\delew$. After a brief review of our naturalness considerations in
Sec.~\ref{sec:nat}, in Sec.~\ref{sec:coupling} we discuss the 
Higgs couplings in the MSSM and in Sec.~\ref{sec:constraints} 
we discuss constraints on natural SUSY parameter space. 
We present in Sec.~\ref{sec:kappa} our main results
of the  values of the $\kappa_i$ versus electroweak naturalness measure
$\delew$ from a scan over parameters of the two-extra parameter
non-universal Higgs (NUHM2) supergravity (SUGRA) model \cite{nuhm2} 
which allows solutions with $\delew$ as low as 5--10. We compare these
expectations against the values which are expected to be probed at
present and future LHC runs, and with expectations from the
International Linear $e^+e^-$ Collider (ILC). In SUSY models with low
$\delew$ (highly natural models), the bulk of points give tiny
deviations from SM expectation. In Sec.~\ref{sec:wino}, we show that the
value of $\kappa_\gamma$ can be enhanced to yield deviations as high as
only 2\% in models with gaugino mass non-universality and a light wino 
\cite{nugm}. In Sec.~\ref{sec:conclude}, we compare our results
against direct sparticle search prospects for LHC and ILC. We stress
there that LHC13 has only a limited reach for natural SUSY. In addition,
if the ILC is built initially as a Higgs factory, we ultimately expect
from natural SUSY that the Higgs profile will look very SM-like: any
major deviation from the SM $\kappa_i$ would likely come from a rather
light spectrum of heavy Higgs bosons which are already highly
constrained by LHC searches. We find that the best prospect for probing
natural supersymmetric models remains as the direct production of
higgsino pairs at ILC. In such a case, ILC would function as a higgsino
factory and as a discovery machine for SUSY\cite{ilc}.

\section{A natural SUSY spectrum}
\label{sec:nat}

Any quantitative discussion of naturalness requires the use of some measure, 
and several have appeared in the literature. Before proceeding, however,
we note the observation that some measures can be mis-applied by
claiming large opposite-sign contributions to observables of {\it
dependent} quantities: these mis-applications lead 
to over-estimates \cite{comp} of
electroweak fine-tuning.\footnote{For example, if an observable ${\cal
O}$ is expressed as ${\cal O}={\cal O}+b -b$ where $b$ is large, then
${\cal O}$ may look fine-tuned. In this trivial example, combining
dependent contributions then cancels the would-be source of fine-tuning.} 

To avoid such pitfalls, any 
naturalness measure should obey the fine-tuning rule \cite{seige}:
{\it When evaluating fine-tuning, it is not permissible to claim fine-tuning of 
{\bf dependent} quantities one against another}.

\subsection{Electroweak scale naturalness}

The relationship between the weak scale $m_{\rm weak}$
and SUSY parameters arises from minimizing the MSSM scalar potential.
One is led to the relation \cite{wss}
\be
\frac{m_Z^2}{2}=\frac{m_{H_d}^2+\Sigma_d^d-(m_{H_u}^2+\Sigma_u^u)\tan^2\beta}{\tan^2\beta -1}-\mu^2\;,
\label{eq:mzs}
\ee
for the $Z$ mass $m_Z$, where $\Sigma_u^u$ and $\Sigma_d^d$ denote the 1-loop
corrections (expressions can be found in the Appendix of
Ref.~\cite{rns}) to the scalar potential, $m_{H_u}^2$ and $m_{H_d}^2$
are the Higgs soft masses, and $\tan \beta \equiv \langle H_u \rangle /
\langle H_d \rangle$ is the ratio of the Higgs VEVs.  
SUSY models requiring large cancellations between the various terms on the
right-hand-side of Eq.~\eqref{eq:mzs} to reproduce the measured value of
$m_Z^2$ are regarded as unnatural, or fine-tuned. In contrast, 
SUSY models which generate terms on the RHS of Eq.~\eqref{eq:mzs} which are all
less than or comparable to $m_{\rm weak}$ are regarded as natural. Thus, the
{\it electroweak} naturalness measure $\delew$ is defined as \cite{rns,mt}
\be
\delew\equiv \text{max}|{\rm each\ additive\ term\ on\ RHS\ of\
Eq.}~\eqref{eq:mzs}| . 
\ee
Including the various radiative corrections, over 40 terms contribute.
Neglecting radiative corrections, and taking moderate-to-large
$\tan\beta \gtrsim 3$, then $m_Z^2/2 \sim-m_{H_u}^2-\mu^2$ so the main
criterion for naturalness is that at the weak scale $m_{H_u}^2\sim
-m_Z^2$ and $\mu^2\sim m_Z^2$ \cite{ccn}. The value of $m_{H_d}^2$
(where $m_A\sim m_{H_d}(\text{weak})$ with $m_A$ being the mass of the CP-odd Higgs
boson) can lie in the TeV range since it is suppressed by $1/\tan^2\beta$.
The largest radiative corrections come from the top squark
sector. Requiring highly mixed TeV-scale top squarks minimizes
$\Sigma_u^u(\tst_{1,2})$ whilst lifting the Higgs mass $m_h$ to $\sim
125$~GeV \cite{rns}. 

Some virtues of $\delew$ include 1. that it is model independent so that
any model generating the same weak scale spectrum will have the same
naturalness value and 2. it obeys the fine-tuning rule.  
It is also predictive: since $|\mu |\sim m_Z$, it implies a spectrum of
four higgsino states $\tz_{1,2}$ and $\tw_1^\pm$ all lying not-too-far
from $m_Z$: $\mu\sim 100-200$ GeV. While many models are indeed highly
fine-tuned under $\delew$\cite{msug,seige}, the NUHM2 model and its
generalizations admit $\delew$ values as low as $5-10$ leading to just
$10-20\%$ electroweak fine-tuning. The models with low $\delew\lesssim
30$ are regarded as {\it natural}. 

\subsection{Higgs mass fine-tuning}

An alternative measure comes from Higgs mass fine-tuning
\cite{Harnik:2003rs,Kitano:2005wc,papucci}. 
The light Higgs mass for $\tan\beta\gtrsim 3$ is given by
\be
m_h^2\sim 
-2 \{m_{H_u}^2(\Lambda )+\delta m_{H_u}^2+\mu^2\}  ~,
\label{eq:ft_hmass}
\ee
where the largest contribution to $\delta m_{H_u}^2$ includes
divergent logarithms of the effective theory cut-off scale $\Lambda$ which
is commonly taken to be $\Lambda\simeq m_{\rm GUT}\sim 2\times 10^{16}$~GeV in
gravity-mediation. By neglecting gauge terms and setting $m_{H_u}^2$ to zero,
a one step integration of the renormalization group equation (RGE) 
for $m_{H_u}^2$ leads to $\delta m_{H_u}^2 \sim
-\frac{3f_t^2}{8\pi^2}(m_{Q_3}^2+m_{U_3}^2+A_t^2)\ln\left(\Lambda/m_{\rm
SUSY} \right) $. 
Requiring $\delta m_{H_u}^2\lesssim m_h^2$ then implies the existence of
three third generation squarks $\tst_{1,2},\ \tb_1$ with mass less than 
about 600~GeV \cite{papucci}. It also leads to claims that SUSY is
fine-tuned at the per-mille level \cite{hall}. 

The problem with this
measure is that $m_{H_u}^2$ and $\delta m_{H_u}^2$ are not
independent.\footnote{This is different from the case of the SM Higgs
mass fine-tuning where the tree level mass and quadratic divergences are
independent.}  
In fact, the larger the value of $m_{H_u}^2(\Lambda )$ becomes, then 
the larger becomes the cancelling correction \cite{arno}. By combining the 
dependent terms $(m_{H_u}^2(\Lambda )+\delta m_{H_u}^2)$, then instead
one  is lead to requiring that $\mu^2$ and  the {\it weak scale value}
of $m_{H_u}^2$ are both comparable to $m_h^2$. Even if
$m_{H_u}^2(\Lambda )$ lies in the multi-TeV range, it can be driven to
small negative squared values at the weak scale $m_{\rm weak}$ via the
same radiative corrections that drive electroweak symmetry breaking in
SUGRA models. 
After combining dependent contributions to $m_h^2$, then a low Higgs mass
fine-tuning implies the same general consequences as those of low
$\delew$. 

\subsection{BG fine-tuning}

The BG measure\cite{eenz,bg,feng} is defined as 
\be
\Delta_{\rm BG}=\text{max}_i\left|\frac{\partial\log m_Z^2}{\partial
\log p_i}\right| ~,
\ee
where $p_i$ are fundamental (usually high scale) parameters of the model
labeled by index $i$. 
To evaluate, we start with the weak scale relation 
\be
m_Z^2\simeq -2\mu^2(\text{weak})-2m_{H_u}^2(\text{weak})~,
\label{eq:mzsapprox}
\ee
and express the right-hand-side in terms of fundamental high scale
parameters. A pitfall can occur in what constitutes high scale
independent parameters. Since $\mu$ hardly evolves during the
renormalization group (RG) flow-- it only receives the wave-function
renormalization thanks to the non-renormalization theorem of the
superpotential-- then $\mu (\text{weak})\sim \mu(\Lambda )$. On the other
hand, $m_{H_u}^2$ evolves greatly: indeed, it must be driven 
through zero to negative  values by the large top quark Yukawa coupling
$f_t$ for electroweak symmetry to be broken radiatively. Semi-analytic
solutions to the $m_{H_u}^2$ RGE allows $m_{H_u}^2({\rm weak})$ to be
evaluated as a large sum of contributions from various 
high scale soft parameters: some positive and some negative. In the case
of gravity-mediation, however, for any particular hidden sector the
high-scale soft terms are all calculable as multiples of the gravitino
mass $m_{3/2}$\cite{sw}. If we vary $m_{3/2}$, the soft terms all vary
accordingly: {\it i.e.} they are {\it not independent} in SUGRA models. 
By combining the dependent soft SUSY breaking terms, 
then the $Z$ mass can be expressed as\cite{seige} 
\be
m_Z^2\simeq -2\mu^2 (\Lambda )-a m_{3/2}^2 ~,
\label{eq:mzsm32}
\ee
with $a$ being a certain proportionality factor dependent on each 
soft mass spectrum. Using Eq.~\eqref{eq:mzsm32}-- and since $\mu$ hardly
evolves from $\Lambda$ to $m_{\rm weak}$-- we have $am_{3/2}^2\simeq 2
m_{H_u}^2 ({\rm weak})$. Even if $m_{3/2}$ is large (as implied by LHC8 limits
for gravity-mediation), then one may still generate natural models if
the coefficient $a$ is small. Under the combination of dependent soft
SUSY breaking terms, then low $\Delta_{\rm BG}$ implies the same as
low $\delew$: that $\mu\sim m_{\rm weak}$ and that $m_{H_u}^2$ is driven to
small and not large negative values.

\section{Higgs couplings in the MSSM}
\label{sec:coupling}

In this Section, we briefly review the Higgs couplings to SM particles.
In the MSSM, the lighter of the two CP-even Higgs mass eigenstates is typically 
a SM-like Higgs boson but with properties differing from the SM case
depending on the mixing angle $\alpha$.
In general, the Higgs boson couplings to vector bosons ($W$ and $Z$ bosons) 
are simply determined by $\alpha$ and $\beta$ while couplings to fermions 
have contributions from loop corrections as well.
On the other hand, the dimension-five couplings of Higgs to
diphoton and to digluon are generated at one-loop order. Note that also
in the SM these couplings are induced at loop level. For this reason,
these couplings can be rather sensitive to the SUSY effects. 

In the MSSM, the SM-like Higgs boson is usually the lighter eigenstate
of the mass matrix 
\be
{\cal M}_h^2=
\begin{pmatrix}
m_{H_u}^2+\mu^2+m_Z^2(1-2\cos 2\beta)/2 & -(m_{H_u}^2+m_{H_d}^2+2\mu^2+m_Z^2)\sin 2\beta/2 \\
-(m_{H_u}^2+m_{H_d}^2+2\mu^2+m_Z^2)\sin 2\beta /2 & m_{H_d}^2+\mu^2+m_Z^2(1+2\cos 2\beta )/2
\end{pmatrix}
+\delta{\cal M}_h^2
\ee
in the basis of the weak eigenstates of the CP-even neutral Higgs
fields $(h^0_{uR},h^0_{dR})$. 
The radiative corrections to the Higgs mass-squared matrix are included in 
$\delta {\cal M}_h^2$.
It is worth noting that we use Higgs soft masses and $\mu$ in order to directly compare it with fine-tuning argument in Eq.~(\ref{eq:ft_hmass}).
The conventional form of mass matrix is obtained if one uses the relations,
\begin{eqnarray}
m_A^2&=&m_{H_u}^2+m_{H_d}^2+2\mu^2\nonumber\\
&=&\frac{1}{\cos^2\beta}\left(m_{H_u}^2+\mu^2-\frac12m_Z^2\cos2\beta\right)\nonumber\\
&=&\frac{1}{\sin^2\beta}\left(m_{H_d}^2+\mu^2+\frac12m_Z^2\cos2\beta\right),
\end{eqnarray}
from the electroweak symmetry breaking conditions.
The mass-squared matrix ${\cal M}_h^2$ is diagonalized by the mixing
matrix\footnote{In the notation of Ref.~\cite{wss}, the Higgs mixing
angle $\alpha$ is the negative of other conventions.} 
\be
\begin{pmatrix}
h \\ H
\end{pmatrix}
=
\begin{pmatrix}
\cos\alpha & \sin\alpha \\
-\sin\alpha & \cos\alpha
\end{pmatrix}
\begin{pmatrix}
h_{uR}^0 \\ h_{dR}^0
\end{pmatrix}
~.
\label{eq:h_mixing}
\ee
The vector boson couplings are simply given by 
\be
g_{hVV}=g_{hVV}^{\rm SM} \sin(\alpha+\beta)\quad\mbox{for}\quad V=W,Z.
\ee
In the decoupling limit where $m_A\gg m_Z$, the mixing angle $\alpha$ follows the relation, $\alpha+\beta\simeq \pi/2$.
This decoupling behavior can be clearly seen in the approximate formula \cite{Carena:2001bg}:
\be
\cos(\alpha+\beta)=\frac{m_Z^2\sin4\beta}{2m_A^2}
\left(
1-\frac{\delta{\cal M}_{h,11}^2-\delta{\cal M}_{h,22}^2}{2m_Z^2\cos2\beta}-\frac{\delta{\cal M}_{h,12}^2}{m_Z^2\sin2\beta}
\right)
+{\cal O}\left(\frac{m_Z^4}{m_A^4}\right).
\label{eq:cosaplb}
\ee
Note that if the radiative corrections in
Eq.~\eqref{eq:cosaplb} are sub-dominant, then $\cos (\alpha + \beta) <
0$ since $\sin 4\beta < 0$. This determines the direction of the
deviation in the Higgs-fermion couplings, as we will discuss below. 
The above equation reads
\be
\sin(\alpha+\beta)\simeq1-\frac12\cos^2(\alpha+\beta)=1-{\cal
O}\left(\frac{m_Z^4}{m_A^4}\right).
\label{eq:sinaplb}
\ee
From this relation, one can easily see that $\cos(\alpha+\beta)\to0$ and
$\sin(\alpha+\beta)\to1$ as $m_A\to \infty$. 
In addition, we note that
the deviations in the gauge boson couplings from the SM ones rapidly
decrease in the large $m_A$ limit since they are suppressed by a factor
of $m_Z^4/m_A^4$. Thus, we expect that the vector boson couplings are
almost SM-like, as will be actually seen in
Sec.~\ref{sec:kappa}. 

For the case of fermion couplings, 
we need to consider an effective Lagrangian\footnote{In our analysis,
the low-energy effective theory is matched  onto the full MSSM at the scale of
$m_{\text{SUSY}}\equiv
(m_{\tilde{t}_1}m_{\tilde{t}_2})^{\frac{1}{2}}$ in the
$\overline{\text{DR}}$-scheme \cite{Siegel:1979wq}, with
$m_{\tilde{t}_{1,2}}$ being the masses of stops. } 
for the Yukawa couplings. 
Including SUSY threshold loop-corrections, the low-energy effective
Yukawa terms below the SUSY breaking scale are written as \cite{Leff,
Pierce:1996zz}
\be
-{\cal L}_{\rm eff}=(f_b+\delta f_b)
\bar{b}_RH_dQ_3+\Delta f_b\bar{b}_RH_u^*Q_3+
(f_t+\delta f_t)\bar{t}_RH_uQ_3+\Delta f_t\bar{t}_RH_d^*Q_3+\mbox{h.c.},
\label{eq:eff_yuk}
\ee
where $\delta f_b$ and $\delta f_t$ represent the radiative corrections
to the tree-level bottom and top Yukawa couplings $f_b$ and $f_t$ in the
MSSM superpotential,
respectively, while $\Delta f_b$ and $\Delta f_t$ are loop-induced 
non-holomorphic Yukawa couplings. Notice that these (non-logarithmic)
radiative corrections are generated by the SUSY breaking effects; in the SUSY
limit, the vertex corrections to the Yukawa couplings vanish
because of the non-renormalization theorem. These
radiative corrections modify the relations between the fermion masses
and the corresponding Yukawa couplings as  
\begin{eqnarray}
m_b&=&f_bv\cos\beta\left(1+\frac{\delta f_b}{f_b} +
\frac{\Delta f_b}{f_b}\tan\beta\right)\equiv
 f_b v \cos\beta(1+\Delta_b)~,\label{eq:mbfb}\\ 
m_t&=&f_tv\sin\beta\left(1+\frac{\delta f_t}{f_t}+
\frac{\Delta f_t}{f_t}\cot\beta\right)\equiv
 f_t v \sin\beta(1+\Delta_t)~, 
\label{eq:mtft}
\end{eqnarray}
where $v\simeq 174$~GeV denotes the Higgs VEV and $\Delta_{b,t}$
are given by 
\begin{eqnarray}
\Delta_b&\simeq&\left[\frac{2\alpha_s}{3\pi}M_3\mu
I(m_{\tilde{b}_1}^2,m_{\tilde{b}_2}^2,M_3^2) 
+\frac{f_t^2}{16\pi^2}\mu A_t I(m_{\tilde{t}_1}^2,m_{\tilde{t}_2}^2,\mu^2)\right]\tan\beta~,\\
\Delta_t&\simeq&- \left[\frac{2\alpha_s}{3\pi}M_3A_t
I(m_{\tilde{t}_1}^2,m_{\tilde{t}_2}^2,M_3^2) 
+\frac{f_b^2}{16\pi^2}\mu^2 I(m_{\tilde{b}_1}^2,m_{\tilde{b}_2}^2,\mu^2)\right]~ .
\end{eqnarray}
Here, $\alpha_s$ denotes the strong gauge coupling constant. $M_3$
and $m_{\tilde{b}_{1,2}}$ are the masses of
gluino and sbottoms, respectively. The loop function is defined by
\be
I(a,b,c)=\frac{ab\ln(a/b)+bc\ln(b/c)+ca\ln(c/a)}{(a-b)(b-c)(a-c)}~.
\ee
This function is of the order of $1/\text{max}\{a^2,b^2,c^2\}$ \cite{Leff}. 
Notice that for large
$\tan\beta$, $\Delta_b \simeq \Delta f_b\tan\beta /f_b$ since the
non-holomorphic correction is enhanced by $\tan\beta$.
The non-holomorphic correction to the top Yukawa coupling is, on the
other hand, suppressed by $\tan\beta$, and thus $\Delta_t \simeq
\delta f_t/f_t$ in this case. 
The modifications in the relations \eqref{eq:mbfb} and \eqref{eq:mtft}, as
well as the deviation from the decoupling limit characterized by
Eq.~\eqref{eq:cosaplb}, change the fermion-Higgs couplings from the SM
ones. We have\footnote{Here, we have used the identities
\begin{eqnarray}
\frac{\sin\alpha}{\cos\beta}&=&\sin(\alpha+\beta)-\tan\beta\cos(\alpha+\beta),\\
\frac{\cos\alpha}{\sin\beta}&=&\sin(\alpha+\beta)+\cot\beta\cos(\alpha+\beta).
\end{eqnarray}
}
\begin{eqnarray}
g_{hbb}&=&g_{hbb}^{\rm SM}\left[\sin(\alpha+\beta)
-\frac{\cos(\alpha+\beta)}{1+\Delta_b}\left\{
\tan\beta -\Delta_b \cot \beta +(\tan\beta + \cot\beta )\frac{\delta f_b}{f_b} 
\right\}\right]~,
\label{eq:g_hbb}\\
g_{htt}&=&g_{htt}^{\rm
 SM}\left[\sin(\alpha+\beta)+\frac{\cos(\alpha+\beta)}{1+\Delta_t} 
\left\{
(1+\Delta_t)\cot \beta -(1+\cot^2\beta)\frac{\Delta f_t}{f_t}
\right\}\right]~.
\end{eqnarray}
Here we express the latter equation in terms of $\Delta_t$ and $\Delta
f_t$ since $\Delta_t \simeq \delta f_t/f_t$. 
One can also obtain a similar relation for the Higgs coupling to tau
lepton, $g_{h\tau\tau}$, by replacing $\Delta_b$ with $\Delta_{\tau}$
and $\delta f_b/f_b$ with $\delta f_\tau/f_\tau$ in Eq.~\eqref{eq:g_hbb}.
In this case 
$\Delta_\tau$ is dominantly induced by the wino-higgsino loop diagram
\cite{Guasch:2001wv}: 
\begin{equation}
 \Delta_\tau \simeq 
-\frac{3\alpha_2}{8\pi}M_2 \mu\tan\beta I(m^2_{\widetilde{\tau}_L},
M^2_2, \mu^2)
~,
\end{equation}
where $\alpha_2$ is the SU(2)$_L$ gauge
coupling constant, $M_2$ is the wino mass, and $m_{\tilde{\tau}_{L}}$ is
the left-handed third generation slepton mass. Here, we have dropped the
sub-dominant bino contribution for brevity. From the
above equations, it is found that the deviation from the SM couplings is
proportional to $m_Z^2/m_A^2$ and therefore becomes quite small in the
large $m_A$ limit. In addition, we see that the deviation in the bottom
and tau couplings is enhanced by $\tan \beta$, while that in the top
coupling is not. Thus, the bottom and tau couplings are more appropriate
to probe the SUSY effects than the top coupling, as will be shown
below. Moreover, we note that as long as the radiative corrections are
moderate, the bottom/tau (top) coupling is always larger (smaller) than
the SM one as $\cos(\alpha+\beta) < 0$. This feature is also found in
the analysis given in Sec.~\ref{sec:kappa}.


For the various Higgs couplings to $gg$ pairs, we use the standard expressions 
including quark and squark triangle diagrams as given in
Ref's~\cite{HHG}. We find that the SUSY effect on the effective gluon
coupling is dominantly given by the stop contribution, which can be
approximately expressed as \cite{Arvanitaki:2011ck} 
\begin{equation}
 \kappa_g \simeq 1+\frac{m_t^2}{4}\left[
\frac{1}{m_{\tilde{t}_1}^2}+\frac{1}{m_{\tilde{t}_2}^2}
-\frac{(A_t-\mu \cot\beta)^2}{m_{\tilde{t}_1}^2m_{\tilde{t}_2}^2}
\right] ~.
\end{equation}
This expression shows that $\kappa_g<1$ occurs only if the stop mixing
is sizable. As discussed above, the RNS scenario with $m_h\simeq 125$ GeV 
requires large stop mixing, 
and thus we expect that the gluon coupling can be smaller than
the SM one in this scenario.

For the Higgs couplings to $\gamma\gamma$, we use standard expressions including
quark, lepton, squark, slepton, $W^\pm$, $H^\pm$ and $\tw_{1,2}^\pm$
loops \cite{HHG}. In the RNS models, higgsinos lie around the
electroweak scale, and thus may give rise to a considerable contribution to
the $\gamma\gamma$ coupling if the higgsinos well mix with winos to have
a sizable coupling with the Higgs boson. This can be achieved when wino
has a relatively small mass. We discuss this possibility in
Sec.~\ref{sec:wino}.

\section{Constraints from low energy observables}
\label{sec:constraints}

We next discuss some low energy observables that constrain the low $m_A$ region in 
our scanned results as shown in the next Section.
As discussed in the previous Section, Higgs couplings to vector bosons and fermions 
suffer deviations mainly from Higgs mixing so that large deviations in the $\kappa_i$ 
are mainly expected when $m_A$ is small.
In such cases, however, loop-mediated contributions to $B$ decays and also 
tree-level contribution mediated by charged Higgs also become larger. In this case, 
$B$-decay observables can constrain this portion of parameter space.

\subsection{$BR(B\to X_s\gamma$)}

For our evaluation, we use the NLO SUSY calculation from Ref.~\cite{bb}
as included in Isatools.
In the MSSM, the two major SUSY contributions come from 
chargino-stop loops and also from the charged Higgs-top loop.
In the large $\tan\beta$ regime, these are approximately given by
\begin{eqnarray}
BR(B\to X_s\gamma)|_{\widetilde{W},\tilde{t}}&\propto& \mu A_t\tan\beta\frac{m_b}{v(1+\Delta_b)}f(m_{\tilde{t}_1},m_{\tilde{t}_2},m_{\widetilde{W}}),\\
BR(B\to X_s\gamma)|_{H^{\pm},t}&\propto&\frac{m_b(f_t\cos\beta-\Delta f_t\sin\beta)}{v\cos\beta(1+\Delta_b)}
g(m_{H^{\pm}},m_t)
\label{eq:bsg_chstop}
\end{eqnarray}
where $f$ and $g$ are loop functions~\cite{Carena:2000uj}.
In order to lift the Higgs mass to 125 GeV, it is normally required that 
stop masses are of order TeV although one of them can be below TeV if 
the maximally-mixed stop scenario is considered.
In such a case, then the chargino-stop loop contributions are usually small.
For the small $m_A$ case-- where Higgs mixing can sizably affect vector boson and 
fermion couplings-- the light charged Higgs ($m_{H_{\pm}}^2\simeq m_A^2+m_W^2$) 
can make a sizable contribution to the decay width. In this case, the 
small $m_A$ region is typically excluded by the $BR(B\to X_s\gamma)$ measurement.

\subsection{$BR(B_s\to \mu^+\mu^-)$}

The $B_s\to \mu^+\mu^-$ decay is induced by flavor changing interactions of Higgs 
bosons, $h$, $H$, and $A$.
The flavor changing couplings of Higgs with $b$- and $s$-quarks are generated by 
similar loop processes as those for $\Delta_b$ and $\Delta_t$.
The physical discussion and calculation details are provided in Ref.~\cite{xt}.
Since $g_{hbs}\propto\cos(\alpha+\beta)/\cos^2\beta$, 
$g_{Hbs}\propto -\sin(\alpha+\beta)/\cos^2\beta$ and $g_{Abs}\propto -1/\cos^2\beta$, 
the dominant contributions are from $H$ and $A$ mediated processes.
The overall branching ratio is then given by
\be
BR(B_s\to \mu^+\mu^-)\propto\tan^6\beta\frac{m_t^4}{m_A^4}.
\label{eq:bsmumu}
\ee
Hence small $m_A$ and large $\tan\beta$ enhance $BR(B_s\to \mu^+\mu^-)$, 
and thus the experimental bounds stringently restrict such
a parameter region. 

%
%
%
%

\section{Results for RNS in NUHM2}
\label{sec:kappa}

In this Section, we explore the $\kappa_i$ values which are expected in
SUGRA GUT models with low fine-tuning $\delew$. We will adopt the
two-extra parameter non-universal Higgs model \cite{nuhm2} (NUHM2) as a
template. The parameters in this model are given by 
\be
m_0,\ m_{1/2},\ A_0,\ \tan\beta,\ \mu,\ m_A .
\ee
In the above set, $m_0$ is the GUT scale value of the common soft mass
parameter for matter scalars, $m_{1/2}$ is the unified gaugino 
mass, $A_0$ is the unified trilinear soft term, $\tan\beta$ is the usual
ratio of VEVs and $\mu$ and $m_A$ are the weak scale values of the
superpotential $\mu$ parameter and the pseudoscalar Higgs mass. These
latter two determine the weak scale values of $m_{H_u}^2$ and
$m_{H_d}^2$ from the scalar potential minimization conditions. The GUT
scale values of $m_{H_u}^2$ and $m_{H_d}^2$ are determined by RG
evolution and are then in general non-universal with the matter scalar
mass $m_0$.  

We generate a random scan over the parameter space
\bea
m_0 &:& \ 0-20\ {\rm TeV}, \nonumber\\
m_{1/2} &:& \  0.5-2\ {\rm TeV},\nonumber\\
-3 &<& A_0/m_0 \ <3,\nonumber\\
\mu &:& \ 0.1-1.5\ {\rm TeV}, \label{eq:param}\\
m_A &:& \ 0.15-20\ {\rm TeV},\nonumber\\
\tan\beta &:& 3-60 , \nonumber
\eea
and generate sparticle mass spectra using Isajet 7.84 \cite{isajet}
which contains the Isasugra subprogram.
The range of $\mu$ covers only positive values; the physical results are
very similar in the case $\mu<0$ except that  
$\Delta$Br$(b\to s \gamma)$ limits are
more constraining for low $m_A$ and hence we get smaller deviations 
in the $\kappa_i$ values.
The major difference with negative $\mu$ is
$\kappa_{\gamma}$ which will be discussed in Sec.~\ref{sec:wino}. 

We require of our solutions that:
\bi
 \item electroweak symmetry be radiatively broken (REWSB),
 \item the neutralino $\tz_1$ is the lightest MSSM particle,
 \item the light chargino mass obeys the model
independent LEP2 limit, $m_{\tw_1}>103.5$~GeV~\cite{lep2ino},
\item LHC search bounds on $m_{\tg}$ and $m_{\tq}$ in mSUGRA are respected\cite{atlas_s,cms_s},
\item $-2.3\times10^{-9}<\Delta $Br$(B_s\to \mu^+\mu^-)<0.6\times10^{-9}$ \cite{Endo:2015oia}
\item $-3.6\times10^{-5}<\Delta $Br$(b\to s \gamma)<9.2\times10^{-5}$ \cite{Endo:2015oia}
\item $m_h=125\pm 2$~GeV.
\ei
Here, we have taken a $\pm 2$ GeV error range in the Higgs mass $m_h$ to reflect
the theoretical uncertainty of the computation.\footnote{We implement the RG-improved 1-loop effective 
potential calculation of $m_h$ which includes leading 2-loop terms in Isajet.} 
The lower bound on the Higgs mass rules out the region where $m_0\lesssim$ 0.4 TeV. 
The upper bound on the branching ratio of $b\to s \gamma$ decay removes the bulk of parameter region 
where $m_A\lesssim$ 0.3 TeV.

\begin{figure}[t]
\begin{center}
 \includegraphics[clip, width = 0.8 \textwidth]{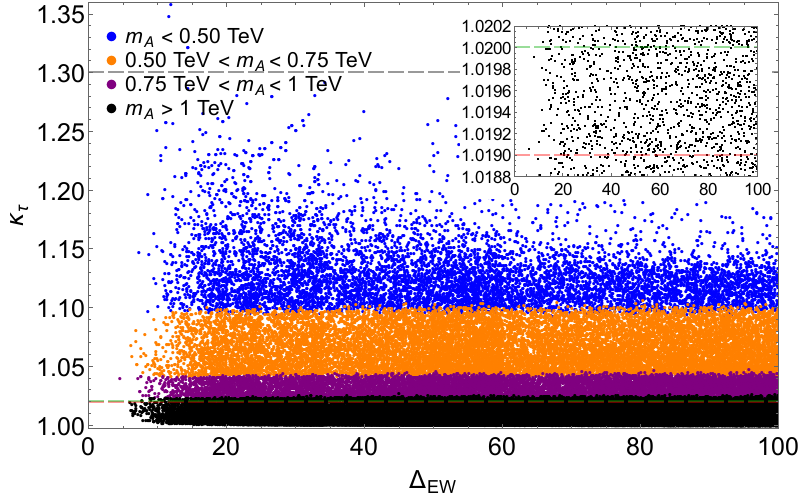}
\caption{$\kappa_\tau$ vs. $\delew$ from scan over NUHM2 parameter
space with $m_h=125\pm 2$~GeV and LHC Higgs and sparticle mass constraints.
The current LHC reach (gray-dashed) and future reach of LHC (green-dashed) and ILC (red-dashed) are shown (see the text for details).
 }
\label{fig:kappa_tau}
\end{center}
\end{figure}
Our first results are shown as $\kappa_\tau$ vs. $\delew$ in
Fig.~\ref{fig:kappa_tau}. Here, the dots are color coded as to the
value of $m_A$, with blue indicating $m_A<0.5$~TeV, orange is
$0.5~\text{TeV}<m_A<0.75$~TeV, purple is $0.75~\text{TeV}<m_A<1$~TeV 
and black is $m_A>1$~TeV.
As discussed in Sec.~\ref{sec:coupling}, this coupling has a large
deviation from unity 
if $\cos(\alpha +\beta)$ is sizable, which occurs when $m_A$ is
comparable to $m_h$. Thus, the magnitude of $\kappa_\tau$ follows
the mass values for the heavy Higgs eigenstates: a value of
$\kappa_\tau\sim 1$ when the heavy Higgs eigenstates decouple. 
Furthermore, in a wide range of parameter space $\kappa_\tau$ is larger
than unity, as discussed in Sec.~\ref{sec:coupling}. In the
RNS model with low $\delew$, $|\mu |\sim m_Z$ and $|m_{H_d}|\sim
m_A$. Since $m_{H_d}^2$ enters $\delew$ as $\sim 
m_{H_d}^2/\tan^2\beta$, then rather large values of $m_A$ are consistent
with low fine-tuning. Upper limits on $m_A$ have been found in
Ref.~\cite{Bae:2014fsa} where $m_A<5-8$~TeV for $\delew<30$ (the exact
upper bound depends on $\tan\beta$). Thus, the bulk of points with
relatively large $m_A$ and low $\delew$ are expected to give only slight
deviations from the SM $h\tau\tau$ coupling. The points with large
deviations occur for low $m_A$, and in fact there are already tight
constraints from LHC on $gg\to h,H,A\to\tau^+\tau^-$ for SUSY in the
$m_A$ vs. $\tan\beta $ plane\cite{Aad:2014vgg}.  
These LHC heavy Higgs search constraints need revision for RNS SUSY
since in RNS the requirement of rather light higgsinos means the heavy
Higgs bosons dominantly decay to charginos and
neutralinos \cite{Bae:2014fsa} rather than to SM modes such as
$\tau^+\tau^-$. We do impose the ATLAS $m_A$ vs. $\tan\beta$ constraints
in the case where $m_{A,H}<2\mu$. 

\begin{figure}[t]
\begin{center}
 \includegraphics[clip, width = 0.8 \textwidth]{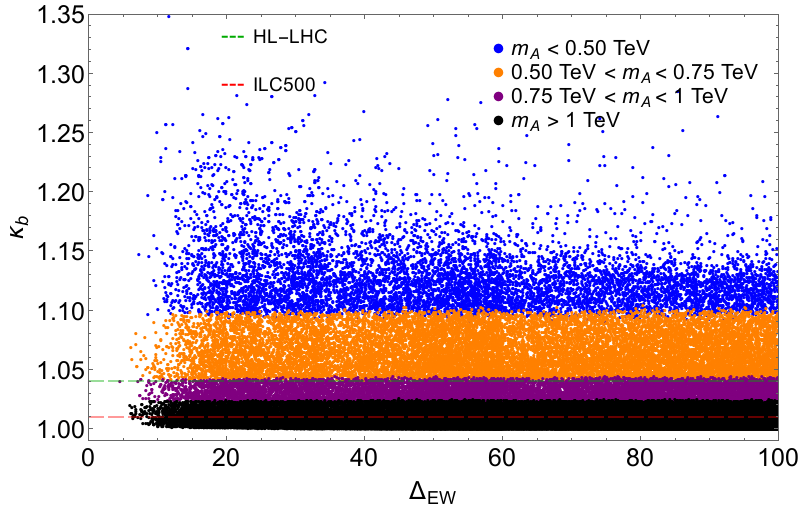}
\caption{$\kappa_b$ vs. $\delew$ from scan over NUHM2 parameter
space with $m_h=125\pm 2$ GeV and LHC Higgs and sparticle mass constraints.
 }
\label{fig:kappa_b}
\end{center}
\end{figure}

In the plot, we also show the current reach for $\kappa_\tau$ from LHC8
as the gray dashed line at $\kappa_\tau\sim 1.3$ and the future reach of high luminosity LHC13
(HL-LHC) with 3000 fb$^{-1}$ and ILC500 in the green- and red-dashed
lines, respectively \cite{Dawson:2013bba, Moortgat-Pick:2015yla}. 
From current reach of LHC8, we can conclude that the LHC experiment has already 
disfavored mass spectra with $m_A<300$~GeV. Furthermore, it turns out that the 
high-luminosity LHC as well as the ILC can probe $m_A\sim 1$~TeV, which can be clearly 
seen in the inset of the figure where a magnified view of $\kappa_\tau$ very
close to 1 is shown. While future colliders can probe much of the
parameter space with low $\delew$, a large chunk with multi-TeV values
of $m_A$ would look very SM-like.

In Fig.~\ref{fig:kappa_b}, we show $\kappa_b$ vs. $\delew$. The locale
of the dots is nearly the same as for the $\kappa_\tau$ plot since
$\kappa_b \simeq\kappa_\tau$. The main difference occurs in the current and future
collider reach for $\kappa_b$. Here, HL-LHC is expected to probe a 4\%
deviation while ILC500 can probe a 1\% deviation. While these reach
values probe a large fraction of parameter space with $m_A\lesssim
1$~TeV, there are a number of natural models which predict quite small
deviation in $\kappa_b$. 

\begin{figure}[t]
\begin{center}
 \includegraphics[clip, width = 0.8 \textwidth]{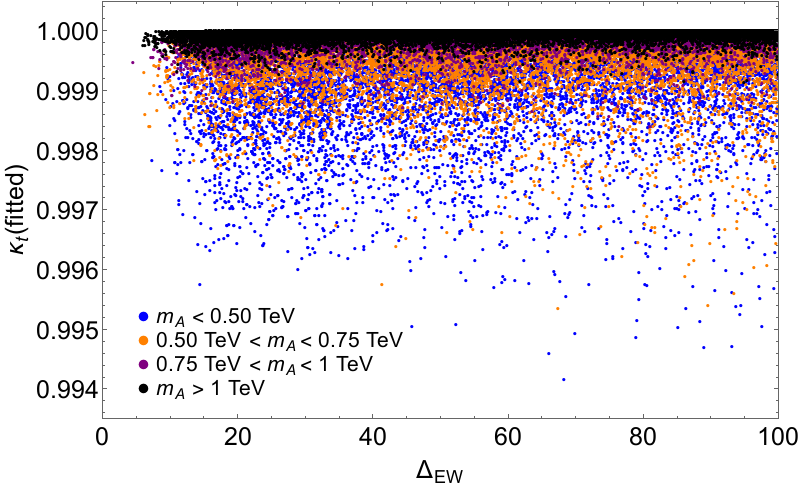}
\caption{$\kappa_t$ vs. $\delew$ from scan over NUHM2 parameter
space with $m_h=125\pm 2$ GeV and LHC Higgs and sparticle mass constraints.
 }
\label{fig:kappa_t}
\end{center}
\end{figure}

In Fig.~\ref{fig:kappa_t}, we show the values of $\kappa_t$ vs. $\delew$. 
As discussed in Sec.~\ref{sec:coupling}, this coupling is expected to
suffer hardly any deviation from the SM value, since there is no
$\tan\beta$ enhanced effect in this case. In addition, the projected 
experimental probes are much more limited since the $h\to t\bar{t}$ decay mode
is kinematically closed. The value of $\kappa_t$ must be extracted from fits to
the Higgs production coupling $hgg$ which includes a top-quark loop in the case of LHC, 
and also to $t\bar{t}h$ production in the case of LHC and ILC. 
Here, it is expected that ILC500 may probe to the 2.5\% level 
($\kappa_t\sim 0.975$) once $\sqrt{s}>2m_t+m_h$.

\begin{figure}[t]
\begin{center}
 \includegraphics[clip, width = 0.8 \textwidth]{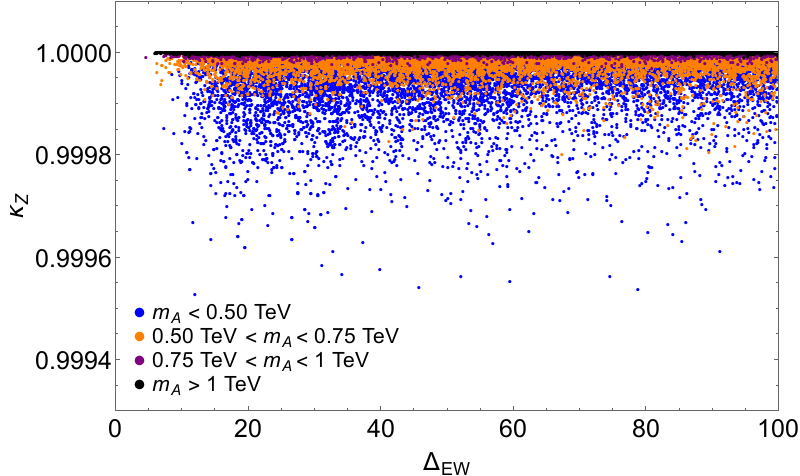}
\caption{$\kappa_Z$ vs. $\delew$ from scan over NUHM2 parameter
space with $m_h=125\pm 2$ GeV and LHC Higgs and sparticle mass constraints.
 }
\label{fig:kappa_z}
\end{center}
\end{figure}

In Fig.~\ref{fig:kappa_z}, we show the values of $\kappa_Z$ vs. $\delew$.
In this case, 
the value of $\kappa_{W,Z}$ is expected to be close to 1 since the
deviation is suppressed by $m_Z^4/m_A^4$ as shown in
Eq.~\eqref{eq:sinaplb}. On the other hand,  
HL-LHC can probe $\kappa_Z$ to $\sim 2\%$ precision via $h\to ZZ^*\to 4\ell$ decays
and ILC can probe to sub-percent precision since $h$ is dominantly produced via
Higgsstrahlung: $e^+e^-\to Z^*\to Zh$. Even so, the bulk of points with 
low $\delew$ have only tiny deviations from 1 and so in this channel
one expects the $h$ to look highly SM-like for RNS SUSY.

\begin{figure}[t]
\begin{center}
 \includegraphics[clip, width = 0.8 \textwidth]{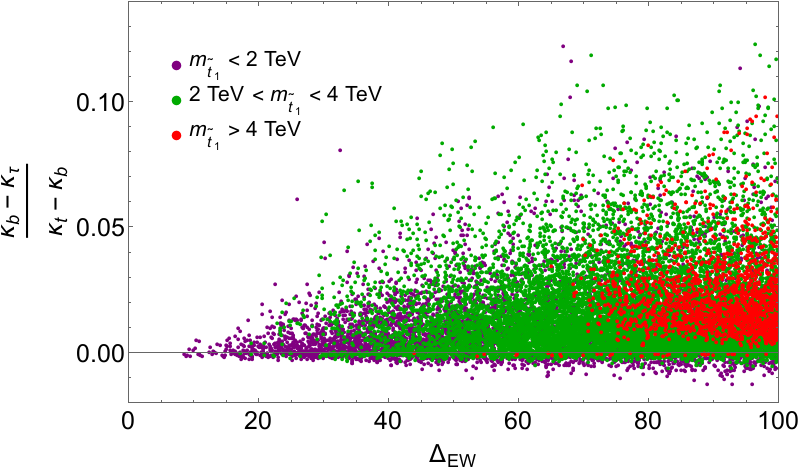}
\caption{$(\kappa_b -\kappa_\tau)/(\kappa_t-\kappa_b)$ vs. $\delew$ from
 scan over NUHM2 parameter space with $m_h=125\pm 2$ GeV and LHC Higgs
 and sparticle mass constraints. 
 }
\label{fig:ratio}
\end{center}
\end{figure}

As we have discussed so far, the deviations in the fermion and gauge
boson couplings are mainly due to the effects of a sizable $\cos (\alpha
+ \beta)$, which occurs if $m_A$ is relatively light. Such effects are,
however, also induced in the two-Higgs doublet models. To confirm the
presence of SUSY effects, therefore, it is desirable to see the
contribution given by other particles than the Higgs bosons. To that
end, we consider the following quantity discussed in
Ref.~\cite{Carena:2001bg}: 
\begin{equation}
 \frac{\kappa_b -\kappa_\tau}{\kappa_t-\kappa_b} \simeq \Delta_b ~,
\end{equation}
where we have kept only the $\tan\beta$-enhanced terms and used the fact
that $|\Delta_\tau|$ is rather small since it is induced by the
electroweak gaugino loop diagrams. In Fig.~\ref{fig:ratio}, we plot
this quantity vs. $\delew$ with color coding in accord with 
$m_{\tst_1}$. It is found that a sizable value of
$\Delta_b$ is expected in most of parameter points. Therefore, we may
extract even the information of the sfermion/gaugino sector via the
precise measurements of the fermion couplings. However, we also note
that the value of the quantity is found to be relatively small in the
small $\delew$ region. This challenges the extraction of $\Delta_b$ in
the natural SUSY scenario.

\begin{figure}[t]
\begin{center}
 \includegraphics[clip, width = 0.8 \textwidth]{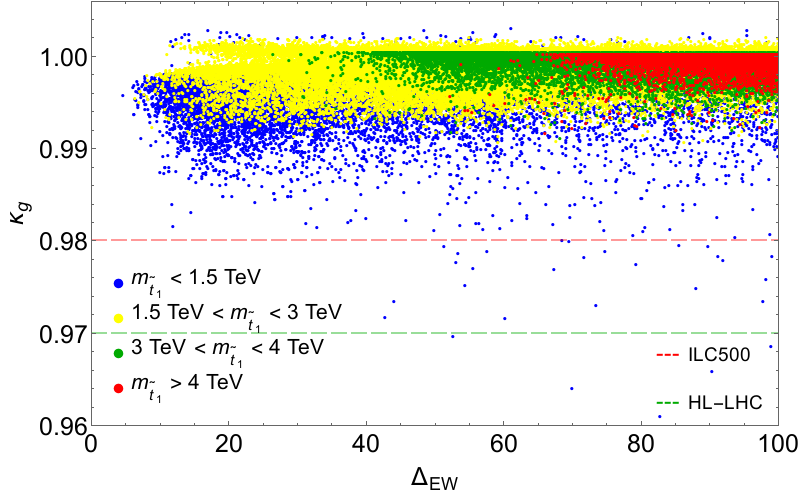}
\caption{$\kappa_g$ vs. $\delew$ from scan over NUHM2 parameter
space with $m_h=125\pm 2$ GeV and LHC Higgs and sparticle mass constraints.
 }
\label{fig:kappa_g}
\end{center}
\end{figure}

In Fig.~\ref{fig:kappa_g}, we show the value of $\kappa_g$ vs. $\delew$.
In this plot, the dots are color coded as to the value of $m_{\tilde t_1}$, 
with blue  indicating $m_{\tilde t_1}<1.5$~TeV, yellow is $1.5$ TeV
$<m_{\tilde t_1}<3$~TeV, green is $3$ TeV$<m_{\tilde t_1}<4$~TeV and 
red is $m_{\tilde t_1}>4$~TeV. Here, the $hgg$ coupling proceeds from triangle diagrams including
quarks for the SM case and also squarks for the SUSY case. Thus, we expect
large deviations from the SM coupling if squarks are far lighter than
the TeV range. Since we require $m_h\sim 125$ GeV, then we implicitly
require TeV-scale highly mixed top squarks which provide a sufficiently
large radiative correction to $m_h$. Usually the top squarks are amongst
the lightest squarks since their masses are suppressed by large
top-quark Yukawa effects in RG running, and also by large
mixing. Furthermore, the NUHM2 model should obey well the LHC8
constraints on the minimal SUGRA model (mSUGRA) \cite{atlas_s,cms_s} so
that $m_{\tq}\gtrsim 1.8$ TeV. 
Thus, we do not expect squarks well below the TeV-scale and therefore
large deviations in the $\kappa_g$ coupling. While some points with low
$m_{\tst_1}$ have deviations of several \%, which can be probed by HL-LHC via
the overall $s$-channel Higgs production rate $\sigma (gg\to h )$, the
bulk of points with a decoupled $m_{\tst_1}$ in the TeV-range tend to have
deviations of less than a percent.
These deviations will be hard to access by either HL-LHC or by ILC.
Note that most of the parameter points predict $\kappa_g <1$. As
discussed above, this can happen only when there exists a large
left-right mixing in the stop mass matrix. Since this large mixing is a
typical feature of the RNS models, the reduction in the $hgg$ coupling
can be regarded as a distinctive prediction in the RNS scenario. 

\begin{figure}[t]
\begin{center}
 \includegraphics[clip, width = 0.8 \textwidth]{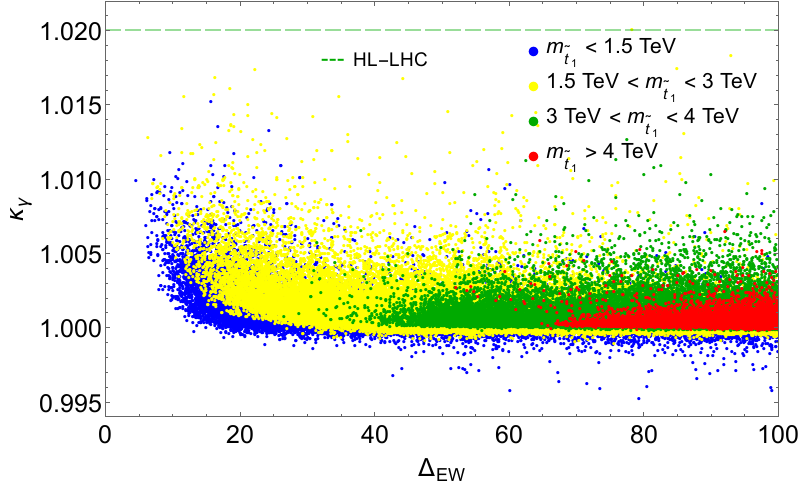}
\caption{$\kappa_\gamma$ vs. $\delew$ from scan over NUHM2 parameter
space with $m_h=125\pm 2$ GeV and LHC Higgs and sparticle mass constraints.
 }
\label{fig:kappa_gam}
\end{center}
\end{figure}

In Fig. \ref{fig:kappa_gam}, we show the value of $\kappa_\gamma$ vs. $\delew$. 
Color coding is the same as in Fig. \ref{fig:kappa_g}.
In the SM, the $\kappa_\gamma$ coupling proceeds via triangle diagrams involving
charged particles which couple to the Higgs: the $q$s, $\ell$s and
$W^\pm$. Among them, top quark and $W$ boson give rise to the dominant
contributions. In the case of SUSY, then there are
additional loops containing squarks, sleptons, charginos and charged
Higgs bosons. 
As in the case of $\kappa_g$, large deviations are obtained in the light stop region which also
coincides with the small $\mu$ region (with light charginos).
Moreover, for $m_{H^\pm}\sim m_A$ small (large Higgs mixing) then
the $hH^+H^-$ coupling can lead to deviations in $\kappa_\gamma$ as in 
two Higgs doublet models. For light charginos, then the $h\tw_1\tw_1$
coupling can be large and also contribute. This coupling is proportional
to higgsino-times-gaugino components of $\tw_1$ and so in the case where
$\tw_1$ is nearly pure higgsino or wino, the coupling is smaller. From
the plot, we expect deviations in $\kappa_\gamma$ $\lesssim 1\%$.

Even though the $h\to\gamma\gamma$ branching fraction is small, the LHC
$gg\to h$ production cross section is large and the $\gamma\gamma$
signature is robust. For comparison, the reach in $\kappa_\gamma$ of
HL-LHC is shown which extends to the $2\%$ level. This is not enough to
access the bulk of low fine-tuned RNS models. The small $h\to
\gamma\gamma$ branching fraction limits the ILC capability to
probe this loop-induced coupling. ILC500 is projected to probe values of
$\kappa_\gamma$ to the 8\% level.\cite{Dawson:2013bba} 

%
\section{Natural SUSY with light wino}
\label{sec:wino}

The results from the previous Section were evaluated in the NUHM2 model
which assumes gaugino mass unification: $M_1=M_2=M_3$ at the GUT scale
so that $M_3\sim 7 M_1$ and $M_2\sim 2M_1$ at the weak scale due to RG evolution.
Then the LHC limit (that $m_{\tg}\gtrsim 1.3$~TeV from the mSUGRA
cascade decay analysis) translates roughly to $M_2\gtrsim 350$~GeV
and $M_1\gtrsim 175$~GeV. This means for RNS SUSY with low $\mu\sim
100-200$~GeV that the light $\tw_1$ which circulates in the
$h\gamma\gamma$ loop is mainly higgsino-like and has somewhat suppressed
couplings. The $h\gamma\gamma$ coupling can be increased in models with
non-universal gaugino masses where $m_{\tg}$ can remain above the LHC8
bound, but now $M_2$ and $M_1$ can be much lower resulting in natural
SUSY with either a wino-like or bino-like LSP \cite{nugm}. 

In the RNS case with non-universal gaugino masses and a lower value of $M_2$, 
then the $\tw_1$ can be a wino-higgsino admixture. Such a mixed chargino
enhances its coupling to the Higgs boson\footnote{See p. 178 of
Ref.~\cite{wss}.} $h$ which depends on a product of gaugino times higgsino
components. 

\begin{figure}[t]
\begin{center}
 \includegraphics[clip, width = 0.7 \textwidth]{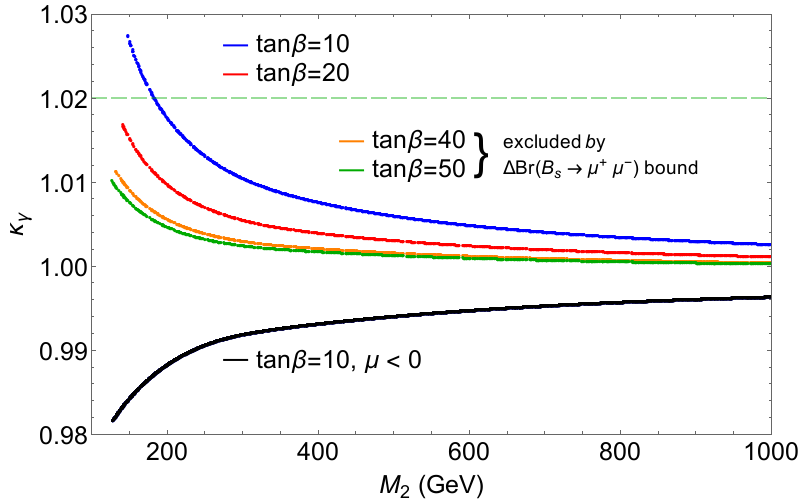}
\caption{$\kappa_\gamma$ vs. $M_2$ along the RNS model line 
for various values of $\tan\beta$. The reach of HL-LHC is shown by the green dashed line.
 }
\label{fig:M2}
\end{center}
\end{figure}

To show a case with maximal $\kappa_\gamma$ in RNS, we plot in Fig. \ref{fig:M2}
the value of $\kappa_\gamma$ vs. $M_2$ along an RNS model-line with parameters
$m_0=5$~TeV, $m_{1/2}=0.7$~TeV, $A_0=-8$~TeV, $\mu=200$~GeV and $m_A=1$~TeV.
We abandon gaugino mass unification and instead allow $M_2$ to vary.
We also plot results for several $\tan\beta$ values. For the case shown, 
as $M_2$ decreases from its universal value, the $\tw_1$ becomes more of a mixed
wino-higgsino state and the coupling $h\tw_1\tw_1$ increases. Correspondingly, 
$\kappa_\gamma$ increases. The maximal $\kappa_\gamma$ reaches $\sim 1.03$ for
lower values of $\tan\beta$ and for $M_2\sim 150$ GeV with $\delew\sim10$.
Such a large value of $\kappa_\gamma$ should be accessible to HL-LHC
as shown by the green dashed line. Larger values of $\tan\beta\gtrsim40$ ($\delew\sim50$) 
are excluded by $B_s\to \mu^+\mu^-$ constraint since 
Br$(B_s\to \mu^+\mu^-)\propto\tan^6\beta$ as stated in Eq.~\eqref{eq:bsmumu}.
%


It is also interesting to see that negative $\mu$ makes $\kappa_{\gamma}$ smaller than unity.
If stops are as heavy as a few TeV, which is required to obtain the 125~GeV Higgs mass, main contributions to Higgs-to-diphoton decay come from chargino loops, so $\kappa_{\gamma}<1$ means that chargino loop contributions destructively interfere the dominant $W$ boson loop contribution.
It is simply understood from the $h\widetilde{W}\widetilde{W}$ coupling,
which is given by\footnote{See Sec.~8.3 and 8.4 of Ref.~\cite{wss} for
complete formulae. 
} 
\begin{eqnarray}
g_{h\widetilde{W}_1\widetilde{W}_1}\simeq g_2~\mbox{sign}(\mu)
\left|
\frac{m_W^2(M_2\cos\beta+\mu\sin\beta)}{M_2^2-\mu^2}
\right|,\\
g_{h\widetilde{W}_2\widetilde{W}_2}\simeq g_2~\mbox{sign}(\mu)
\left|
\frac{m_W^2(M_2\sin\beta+\mu\cos\beta)}{M_2^2-\mu^2}
\right|,
\end{eqnarray}
where $m_A^2\gg m_h^2$, $\left|M_2^2-\mu^2\right|\gg m_W^2$ and $\left|M_2/\mu\right|<\tan\beta$.
Here we assume that $\widetilde{W}_1$ is mostly higgsino-like and $\widetilde{W}_2$ is mostly wino-like.
The chargino-Higgs couplings flip their sign when the sign of $\mu$ is flipped, and thus chargino contributions can be either
constructive and destructive depending on the sign of $\mu$. In order to avoid chargino LSP, for $\mu<0$ we set $M_1=100$ GeV.
$\kappa_{\gamma}$ can show about 2\% deficit for small $\tan\beta$ and
$M_2$, and it approaches to the SM value as $M_2$ increases (black curve).

If a deviation in $\kappa_\gamma$ is actually observed at the LHC, this
may indicate the presence of a light chargino with sizable coupling to the
Higgs boson. Such a light chargino should be within the reach of direct
production at the ILC. Moreover, a large coupling to the Higgs boson
implies a large neutralino-nucleon scattering cross section. Although we
expect that only a small portion of dark matter energy density is
occupied by the $\widetilde{Z}_1$ LSP since such a light higgsino-like
neutralino in general results in a small relic abundance, it is found
that future dark matter direct detection experiments can probe the
$\widetilde{Z}_1$ LSP in this case, and thus provide a way of examining
RNS models\cite{rnsdm}.

\section{Conclusions:} 
\label{sec:conclude}

In this paper we have presented expectations for possible deviations in
Higgs couplings that are expected in SUSY models with radiatively-driven
naturalness. Such models with low $\delew\lesssim 30$ are natural in the
electroweak sector and, if augmented with a Peccei-Quinn sector, are
natural in the QCD sector as well. Models with a SUSY DFSZ axion also
admit an elegant solution to the SUSY $\mu$ problem. Such natural SUSY
models are consistent with squark, slepton and gravitino masses in the
multi-TeV range which admits a solution to the gravitino problem
\cite{Weinberg:1982zq, Kawasaki:2008qe} and at least a partial
decoupling solution to the SUSY flavor and CP problems
\cite{Gabbiani:1996hi}. These models are rather simple extensions of the
SM and may even be regarded as more conservative than the SM
in that they contain solutions to the gauge hierarchy and strong CP problems. 
Thus, every avenue for their verification must be explored.
Here, we examined the case of Higgs boson profiling.

Our results may be summarized as follows.
\bi
\item Substantial deviations in $\kappa_\tau$ and $\kappa_b$ may be expected for
RNS SUSY but mainly in the case where $m_A$ is rather light leading to
significant mixing in the scalar Higgs sector. However, since $m_A$ can
extend into the multi-TeV range at little cost to naturalness (due to
$\tan^2\beta$ suppression of the term including $m_{H_d}^2$ in Eq.~\eqref{eq:mzs}) these
deviations may well lie below the reach of HL-LHC and even ILC500. 

\item Tiny deviations to $\kappa_t$ are expected. This coupling is also
difficult to measure unless one has a linear $e^+e^-$ collider with
$\sqrt{s}>m_h+2m_t$. 

\item Tiny deviations are expected in $\kappa_{W,Z}$, usually below the
0.5\% level. 

\item Some deviations can occur in the $\kappa_g$ coupling, but mainly
for anomalous cases with very light top squarks $\tst_1$.
However, light top squarks generally lead to large deviations in 
$BF(b\to s\gamma )$ and also have recently been tightly constrained by 
top-squark pair production searches at LHC8\cite{Aad:2014kra, CMS:2014wsa}. 
Except for such cases, most RNS predictions for $\kappa_g$ 
lie below the reach of HL-LHC and ILC500. 

\item Small deviations in $\kappa_\gamma$ are expected-- usually at the
sub-$0.5\%$ level-- below the reach of HL-LHC and ILC500. However, in
models with non-universal gaugino masses where the light chargino
becomes a wino-higgsino mixture, then $\kappa_\gamma$ may increase to
the 1--3\% level. 
\ei  

To summarize: except for unusual cases (highly mixed Higgs sector with
low $m_A$, anomalously light stops soon to be excluded by LHC or highly
mixed charginos) natural SUSY predicts minimal deviations from a SM-like
portrait of the light Higgs boson. Given this situation, it is useful to
compare these indirect search methods against the direct search for
natural SUSY at LHC and ILC. 
\begin{figure}
\hfill
\subfigure[]
{\includegraphics[width=8.cm]{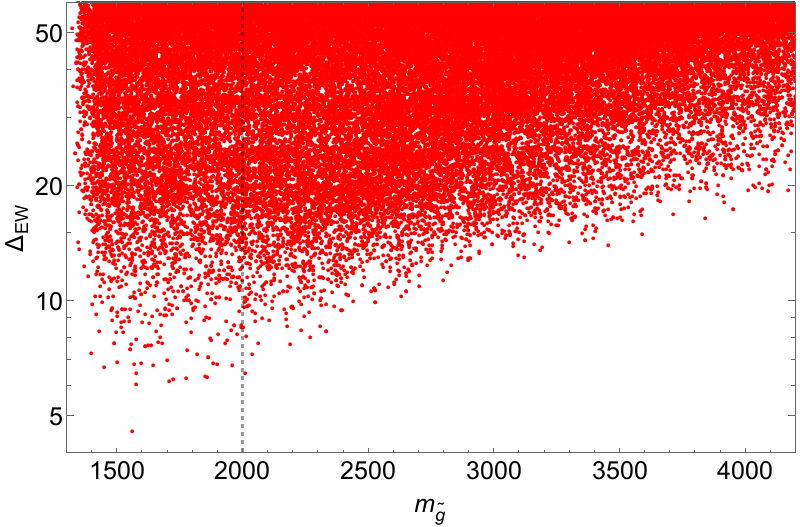}}
\hfill
\subfigure[]
{\includegraphics[width=8.2cm]{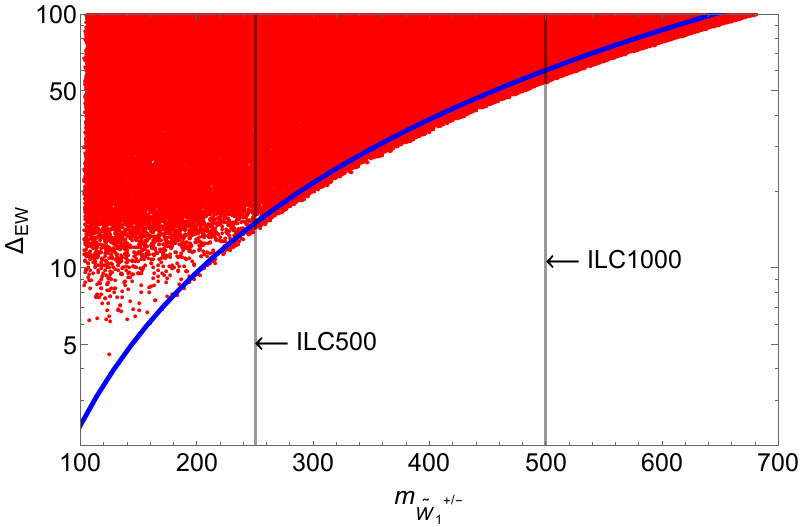}}
\hfill
\caption{(a) Plot of $\delew$ vs. $m_{\tilde g}$ from a scan over NUHM2 
parameter space. (b) Plot of $\delew$ vs. $m_{\tw_1}$ from a scan over NUHM2 
parameter space. We also show the curve $\delew=\mu^2/(m_Z^2/2)$ 
and the reach of various ILC energy options for higgsino pair
production.}
\label{fig:ilc}
\end{figure}

\begin{figure}[t]
\begin{center}
 \includegraphics[clip, width = 0.9 \textwidth]{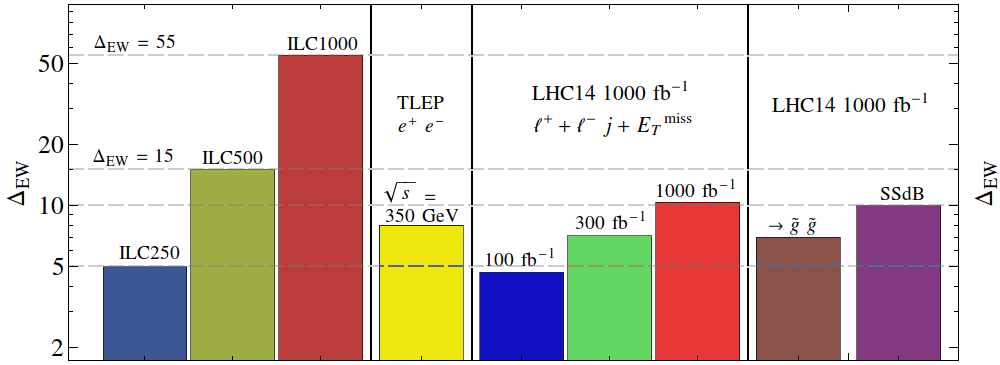}
\caption{Plot of the reach of various LHC and ILC options for
SUSY with radiatively-driven naturalness in terms of
$\delew$.
 }
\label{fig:bar}
\end{center}
\end{figure}

A direct search for $\tg\tg$ production at LHC13 with 1000 fb$^{-1}$
can reach up to $m_{\tg}\sim 2$~TeV \cite{lhc}. This approximately covers $\delew<7$ 
as seen in Fig.~\ref{fig:ilc}(a). 
The LHC13 1000 fb$^{-1}$ reach for $\tg\tg$ production is
also shown in terms of $\delew$ by the brown histogram of Fig.~\ref{fig:bar}. 
LHC13 can also search for light higgsino pair production $pp\to \tz_1\tz_2$ 
where $\tz_2\to\mu^+\mu^-\tz_1$. Since the dimuons tend to be rather
soft (due to the small $m_{\tz_2}-m_{\tz_1}$ mass gap) a trigger on hard
jet radiation from the initial state is needed\cite{jll}. The reach of various
LHC13 options for $\mu^+\mu^- j+\eslt$ production is also shown 
in Fig.~\ref{fig:bar}. 

The most direct test of SUSY naturalness occurs via the direct search for 
higgsino pair production at an $e^+e^-$ collider with $\sqrt{s}>2\mu$.
Such a machine would be a higgsino factory\cite{ilc} in addition to a Higgs factory. 
The value of $m_{\tw_1}$ is plotted versus $\delew$ in 
Fig. \ref{fig:ilc}(b) which exhibits the tight correlation where $\sqrt{s}>2m_{\tw_1}$ 
and where $m_{\tw_1}\simeq \mu$. Since $\delew\sim \mu^2/(m_Z^2/2)$, then
ILC probes directly values of $\delew$ according to
$\sqrt{s}\sim \sqrt{2\delew}m_Z$. From the plot, we see that 
ILC500 makes a complete probe of  $\delew<15$ and ILC1000 probes
$\delew <55$.

In Fig. \ref{fig:bar}, we show the reach in $\delew$ of prospective experiments. 
ILC1000 can see the entire RNS parameter space whereas LHC14 and TLEP 
can probe only a portion of it.
Light higgsinos should ultimately be detected at ILC with $\sqrt{s}>2\mu$. 

\section*{Acknowledgments}

This work was supported in part by the US Department of Energy, Office of High
Energy Physics. The work of NN is supported by Research Fellowships of
the Japan Society for the Promotion of Science for Young
Scientists. KJB, HB and HS  would like to thank the William I. Fine
Institute for Theoretical Physics (FTPI) at the University of Minnesota
for hospitality while this work was initiated. The computing for this project was 
performed at the OU Supercomputing Center for Education \& Research (OSCER) at the 
University of Oklahoma (OU).

%

%
\end{document}